\documentclass[twocolumn,preprintnumbers,amsmath,amssymb]{revtex4}

\usepackage{graphicx}

\bibpunct{}{}{,}{s}{,}{,}

\begin{document}

\preprint{Preprint Dated \today}

\title{\large \bf{Screening of Excitons in Single,\\Suspended Carbon Nanotubes}}

\author{Andrew G. Walsh\footnotemark[2]}
\author{A. Nickolas Vamivakas\footnotemark[4]}
\author{Yan Yin\footnotemark[2]}
\author{Stephen B. Cronin\footnotemark[5]}
\author{M. Selim \"{U}nl\"{u}\footnotemark[4]\footnotemark[2]}
\author{\\Bennett B. Goldberg\footnotemark[2]\footnotemark[4]}
\author{Anna K. Swan\footnotemark[1]\footnotemark[4]}
\affiliation{\fontfamily{ptm}\selectfont Department of Physics, Boston University, 590 Commonwealth Avenue, and Department of Electrical and Computer Engineering, Boston University, 8 Saint Mary's Street, Boston, MA 02215\\Department of Electrical Engineering, University of Southern California, Powell Hall of Engineering PHE 624, Los Angeles, CA 90089-0271}

\begin{abstract}

\begin{center}\bf{ABSTRACT}\end{center}

\bf{\noindent Resonant Raman spectroscopy of single carbon nanotubes suspended across trenches displays red shifts of up to 30 meV of the electronic transition energies as a function of the surrounding dielectric environment.  We develop a simple scaling relationship between the exciton binding energy and the external dielectric function and thus quantify the effect of screening.  Our results imply that the underlying particle interaction energies change by hundreds of meV.}

\end{abstract}

\maketitle

\fontfamily{ptm}\selectfont

\footnotetext[1]{Corresponding author. E-mail: swan@bu.edu.}
\footnotetext[2]{Department of Physics, BU.}
\footnotetext[4]{Department of Electrical and Computer Engineering, BU.}
\footnotetext[5]{Department of Electrical Engineering, USC.}

\noindent The long predicted presence of excitons with large binding energies in carbon nanotubes (CNT)\cite{Ando,Kane1,Kane2,Spataru,Perebeinos} has been experimentally confirmed by recent two-photon experiments \cite{Wang2,Maultzsch1,Dukovic}.  With binding energies of hundreds of meV and Coulomb energies highly sensitive to screening due to the one dimensional nature of CNTs, one expects that the measured optical transition energies should change significantly with changes in the dielectric environment.  Yet experiments report variations on a scale of just a few tens of meV across dielectric environments as different as CNT bundles in solution, micelle encapsulated CNTs, and individual nanotubes suspended in air \cite{Moore,Lefebvre1}.  Lefebvre \textit{et al.} measured the photoluminescence (PL) emission from CNTs freely suspended in air \cite{Lefebvre1} and compared the results with the PL from micelle encapsulated nanotubes published by Bachilo \textit{et al.} \cite{Bachilo}.  By using family structure to correlate CNT species between the two data sets, they were able to show average red shifts of only 28 meV and 16 meV in E$_{11}$ and E$_{22}$, respectively (where E$_{ii}$ is the optical transition energy associated with the \textit{i}$^{th}$ subband), upon micelle encapsulation, a surprisingly small change given the different environments.

In this work, we investigate the underlying reasons for this small variation of the observed optical transition energies.  We follow the shift of the electronic energy levels as we control the screening of the Coulomb interaction in single CNTs suspended across trenches.  Specifically, we use resonant Raman spectroscopy (RRS) to probe the optical transition energy E$_{22}$ of a given CNT as we change the dielectric environment from dry N$_{2}$ to high humidity N$_{2}$ to water.  We present experimental evidence of dramatic underlying changes in those particle interaction energies that largely cancel each other, leading to the small variations in observed optical transition energies, in accordance with the picture described by Ando and Kane and Mele \cite{Ando,Kane1}.

Typical spectra from resonant Raman scattering of single CVD grown CNTs suspended across trenches\cite{SingleCNTs} are shown in
\begin{figure}
\begin{center}
\includegraphics[width=3.3in]{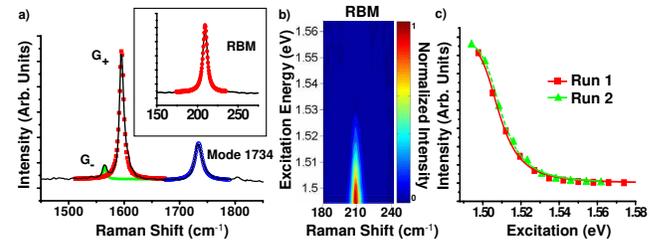}
\caption{(a) Typical Raman spectrum taken in dry N$_{2}$ at 742 nm excitation, near outgoing resonance of the G band.  Lorentzian fits to the G$_{-}$, G$_{+}$ and 1734 cm$^{-1}$ Raman modes are shown in green, red, and blue respectively. \textbf{Inset.} Lorentzian fit, shown in red, of the RBM Raman mode taken at 828 nm, near incoming resonance. (b) Contour plot showing intensity of the RBM as a function of excitation energy and Raman shift.  (c) REPs of the same data set as in (b).}
\label{fig1}
\end{center}
\end{figure}
Figure~\ref{fig1}a.  Details of the experiment have been presented elsewhere. \cite{Yin1, Yin2}  The Stokes scattering peak intensities for each Raman active phonon mode appearing in the spectra are plotted against laser excitation energy resulting in the resonance excitation profile (REP) for a given phonon mode.  In general, the REP will be double peaked from the combined resonances of the incoming and outgoing photons with the electronic structure of the nanotube, which are resolvable when the scattered phonon energy is greater than the broadening of the resonance.  The radial breathing mode (RBM) appears as a single peak whereas the peaks of the incoming and outgoing resonances of the tangential phonon modes, G$_{-}$ and G$_{+}$, are spectrally separated by virtue of their much larger phonon mode energies.  By fitting a one phonon exciton-mediated REP line shape \cite{Vamivakas}, we can determine the electronic transition energy, E$_{ii}$, with which the photons are resonant.  Here we use E$_{ii}$ to denote the excitonic transition associated with the \textit{i}$\rightarrow$\textit{i} interband transition.  Together with the RBM, E$_{ii}$ determines the CNT species.  The data shown in Figures \ref{fig1} and \ref{fig2} are ascribed to a (12,4) nanotube.

\begin{figure}
\begin{center}
\includegraphics[width=3.3in]{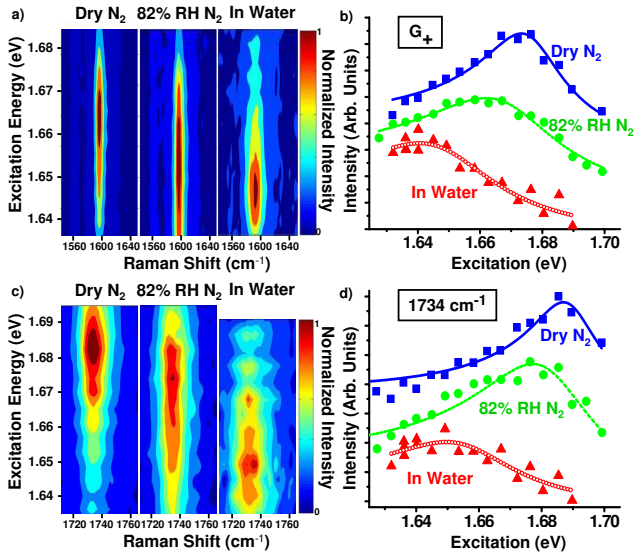}
\caption{(a) Contour plot showing normalized intensity of the G$_{+}$ outgoing resonance as a function of both excitation energy and Raman shift measured in dry N$_{2}$, 82\% RH N$_{2}$, and immersed in water.  (b) REPs of the data shown in (a). Plots are offset in the vertical direction for clarity.  Data measured in dry N$_{2}$ is shown in blue squares with the accompanying REP fit shown with a solid blue line.  Data measured in 82\% RH N$_{2}$ and in water are shown similarly in green and red, respectively.  (c) and (d) Same as in (a) and (b), respectively, but for the 1734 cm$^{-1}$ Raman mode.}
\label{fig2}
\end{center}
\end{figure}

We measure REPs of individually resonant CNTs in dry N$_{2}$ in an enclosed chamber, followed by adding water vapor to the nitrogen with humidity measured with a hygrometer.  Finally, the sample is directly immersed in water and the experiment repeated a third time, all on the same singly resonant CNT.  The set of REPs yield E$_{ii}$ for each phonon mode and dielectric environment, and thus measure the shift in the electronic level with increasing $\epsilon$.  The shifts measured by each phonon mode are all identical for the same CNT, as they should be.

Single nanotube RRS is complicated by the need for high precision positioning and high stability ($\leq$25nm) during spectral acquisition and changing laser frequencies, all at low powers to maintain the CNT phonon bath at room temperature.  Consistency in the measurements is demonstrated in Figures \ref{fig1}b and \ref{fig1}c where the REP of the RBM taken in dry N$_{2}$ is shown.  The electronic resonance, E$_{22}$, falls just outside our experimentally accessible range for this mode.  Despite having only half of the REP for this Raman mode, run 2, which was taken immediately after run 1, predicts the same E$_{22}$ as run 1 to within 1 meV.  In addition, the outgoing peak of the REP for the G$_{+}$ tangential phonon mode for this CNT, shown in Figures \ref{fig2}a and \ref{fig2}b (data taken concurrently with the data shown in Figure\ref{fig1}), yields the same E$_{22}$ as the RBM to within 2 meV.  As expected, all modes yield the same E$_{22}$ for this CNT in dry N$_{2}$, E$_{22}$=1.475$\pm$3 meV.

The results of increasing external dielectric environment on the E$_{ii}$ for two different Raman modes from the same CNT are shown in Figure \ref{fig2}.  The data clearly shows the outgoing peak of the REP red shifting with increasing external dielectric for both the G$_{+}$ and 1734 cm$^{-1}$ Raman modes\cite{Brar}.  The measurements in water are noisier due to lower signal level as a result of perturbation of the wave fronts by the water-coverslip and coverslip-air interfaces.  Despite this, shifts in the resonance peak energies are clearly visible.  High relative humidity N$_{2}$ introduces a $\sim$10 meV red shift in E$_{22}$.  Liquid water red shifts E$_{22}$ by $\sim$30 meV, similar to the differences reported by Lefebvre \textit{et al.}\cite{Lefebvre1}  Table \ref{tab1} shows tabulated results for two different nanotubes.  Note the consistency in the observed shifts between each dielectric environment.  

\begin{table}
\begin{center}
\includegraphics[width=3.3in]{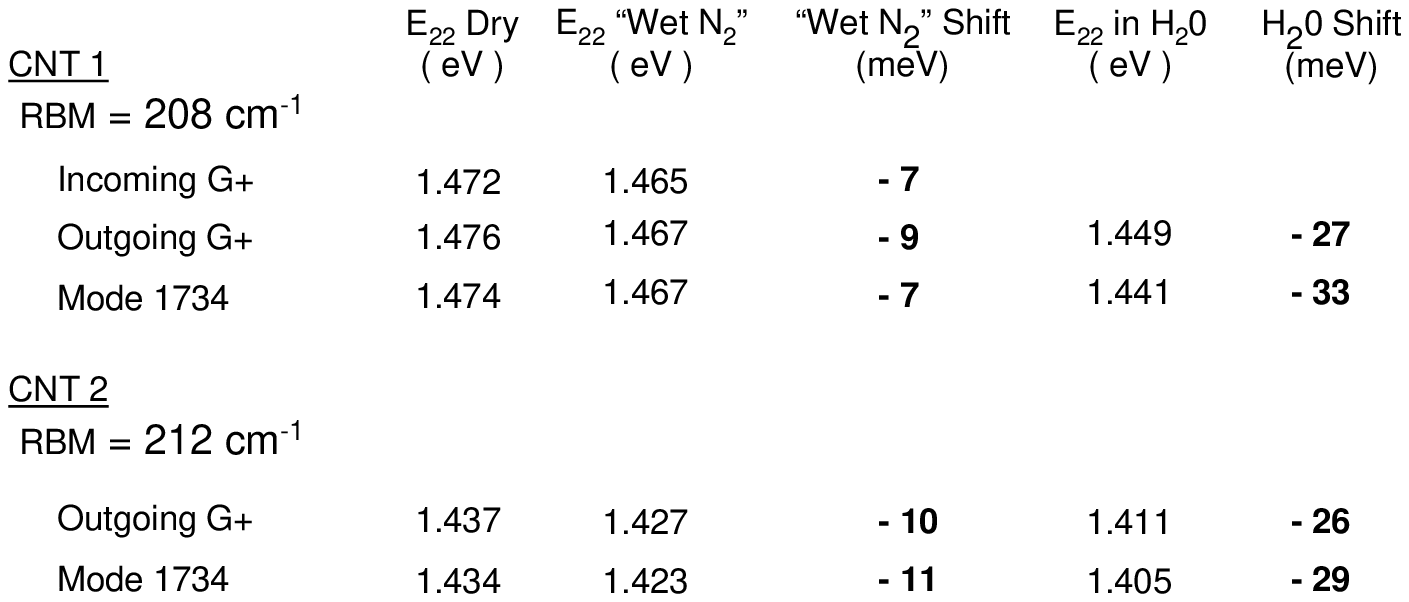}
\caption{Tabulated results of red shifts for 2 carbon nanotubes.  E$_{22}$, as determined by REP fit, is shown for each phonon mode and resonance condition.}
\label{tab1}
\end{center}
\end{table}

In order to use our results to quantify the effect of screening on the particle interaction energies in CNTs, we first must discount other possible environmental influences on the electronic transition energies including temperature\cite{Cronin1,Capaz1,Lefebvre2}, mechanical strain \cite{Cronin1,Cronin2,Cronin3}, and charge transfer \cite{Shim1,Zahab,Collins,Lee,Kong, Shim2,Rao,Claye}.  All measurements were taken at room temperature and laser power kept sufficiently low to avoid heating of the CNT \cite{Cronin1}.  Mechanical strain changes the C-C bond lengths which shifts the energy levels of the nanotube \cite{Cronin1,Capaz2} and can lead to E$_{ii}$ shifts depending on the type of strain (uniaxial, isotropic, radial), (n-m)mod3 value, sub-band index, and chiral angle.  But strain causes a change in the observed phonon energies \cite{Cronin1,Cronin2,Cronin3} and we observe no changes in any phonon mode energies (to within $\sim$ 1 cm$^{-1}$).  A number of studies have investigated charge transfer and its effect on transport\cite{Shim1,Zahab,Collins,Lee,Kong} but it is difficult to separate the effect of charging from that of screening in such experiments.  However, charge transfer has been shown to be associated with a change in the tangential phonon energy\cite{Shim2,Rao,Claye} and since, again, we observe no changes in any phonon mode energies, we believe that charge transfer is negligible, in agreement the diameter dependent activation model proposed by Shim \textit{et al.} \cite{Shim2}.  Thus the primary mechanism for the observed shifting of the electronic transitions is screening of the particle interactions.

\begin{figure}
\begin{center}
\includegraphics[width=3.3in]{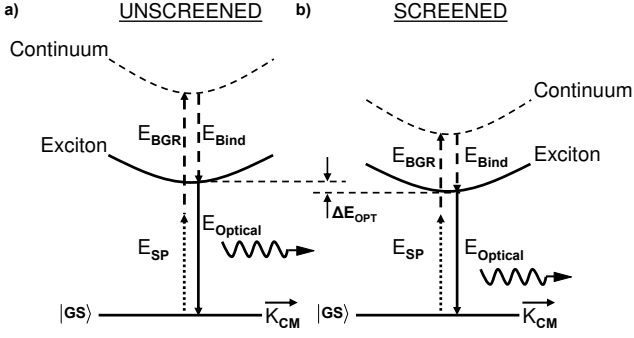}
\caption{Energy diagrams of the effect of the band gap renormalization and exciton binding energies on the optical transition energy in (a) unscreened and (b) screened environments.}
\label{fig3}
\end{center}
\end{figure}

To model the effect of the changing external dielectric function on the particle interaction energies, the simple single particle picture must be modified.  The electron-electron interaction energy, E$_{BGR}$, significantly increases the single particle band gap, largely counteracting the effect of the strong exciton binding energy, E$_{Bind}$, on the optical transition energy \cite{Ando, Kane1}, as depicted schematically in Figure~\ref{fig3}a.  This effect occurs for all sub-bands E$_{ii}$.

We consider the single particle Hamiltonian with two particle interaction terms representing electron-electron and electron-hole interactions, respectively, to determine the optical transition energy measured in our experiment, i.e. E$_{Opt}$ = E$_{SP}$ + E$_{BGR}$ - E$_{Bind}$.  We seek to derive how those two terms scale with changing external dielectric.  E$_{BGR}$ should scale simply as $\epsilon^{-1}$ for small electron wave vectors near the zone center \cite{Kane1}.  Determining the scaling of the exciton binding energy with external dielectric is more difficult.  Most theoretical work treats the dielectric function as a constant fitting parameter and, moreover, does not address the issue of the presence of two different dielectric materials, i.e. the environment and the nanotube itself.  Two works do, however, explicitly address excitonic binding in quantum wires of one dielectric, $\epsilon_{1}$, in an external dielectric $\epsilon_{2}$ \cite{Ogawa2,Banyai} and serve as the starting point of our model.  In both papers, the Coulomb interaction is integrated over the lateral spatial dimensions leading to expressions for the axial, 1D effective potential.  This potential, V$_{eff}^{1D}$(z), where z is the electron-hole separation, can be evaluated numerically and, for the range of dielectric values of interest here, fit very well by a 1/($\vert$z$\vert$+z$_{0}$) potential where z$_{0}$ is known as the cutoff parameter and removes the singularity at zero separation.  We find, for both models, over our range of dielectric values, that z$_{0}$ scales linearly with 1/$\epsilon_{2}$.  This is useful since the exciton binding energy for this potential has been solved \cite{Loundon}.  Specifically, the exciton binding energy is given by E$_{Bind}$ = R$^{*}_{h}/\lambda^{2}$, where the quantum number $\lambda$ is a complicated function of z$_{0}$, R$^{*}_{h}$ is the effective Rydberg defined by R$^{*}_{h}$ = $\mu$e$^{4}$/2$\hbar^{2}\epsilon^{2}$, $\mu$ is the exciton effective mass, and $\epsilon$ is the dielectric constant which is, again, a poorly defined quantity in a heterogeneous environment.  We assume that the dependence of the effective Rydberg on $\epsilon_{2}$ is approximately the same, that is R$^{*}_{h}$ $\propto$ 1/$\epsilon_{2}^{2}$.  The expression for $\lambda$ as a function of z$_{0}$ has been solved numerically \cite{Combescot} and may be approximated by a simple $\lambda$ $\propto$ z$_{0}^{\beta}$ power law where $\beta$$\sim$0.4.

Hence, by combining the scaling relationships between R$^{*}_{h}$ and $\epsilon_{2}$, $\lambda$ and z$_{0}$, and z$_{0}$ and $\epsilon_{2}$ we derive a scaling relationship between the exciton binding energy and $\epsilon_{2}$.  We find that the exciton binding energy should scale as E$_{Bind}$ $\propto$ (1/$\epsilon_{2}^{2}$)/(1/$\epsilon_{2}^{2\beta}$) or $\epsilon_{2}^{-\alpha}$ where $\alpha$=1.2.  This is close to the value 1.4 derived by Perebeinos \textit{et al.} \cite{Perebeinos} for higher dielectric environment where the heterogeneous nature of the dielectric environment was not considered.  Further, since z$_{0}$ scales with the radius of the nanotube \cite{Ogawa1,Ogawa2} and the effective Rydberg is independent of radius, the exciton binding energy therefore scales as E$_{Bind}$ $\propto$ 1/r$^{2*\beta}$ or 1/r$^{0.8}$, close to the previously published result 1/r$^{0.6}$ \ \cite{Perebeinos,Pedersen1}.

\begin{table}
\begin{center}
\includegraphics[width=3.3in]{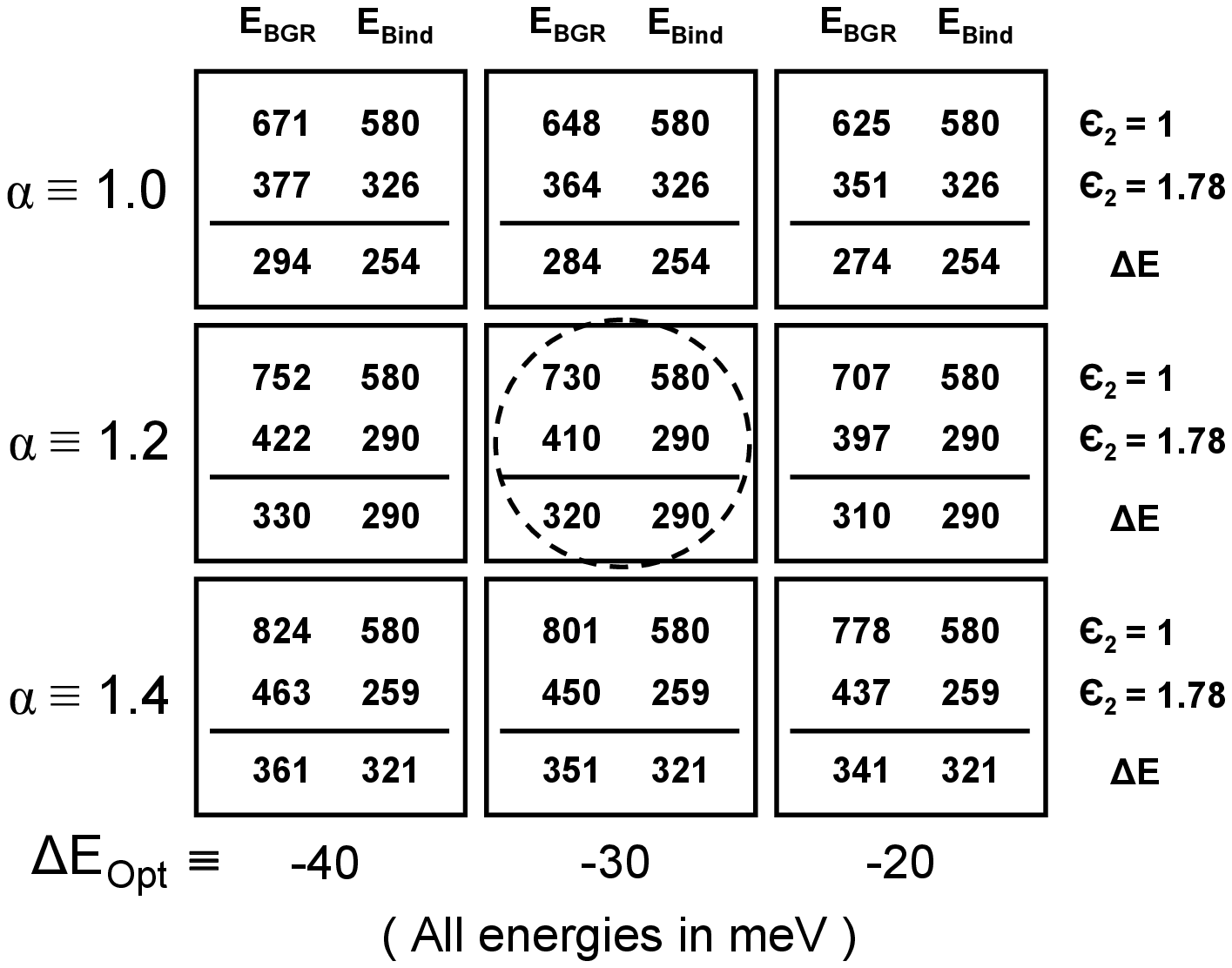}
\caption{Particle interaction energies as a function of the exciton scaling exponent, $\alpha$, the external dielectric, $\epsilon_{2}$, and the change in the optical transition energy, $\Delta$E$_{Opt}$.  An unscreened exciton binding energy of 580 meV \cite{Capaz3} is used as an input parameter.  The values below the line in each box show the changes of the BGR and exciton binding energies with screening.  Their difference is  $\Delta$E$_{Opt}$.  The dashed circle highlights the numbers quoted in the text.}
\label{tab2}
\end{center}
\end{table}

We can now use these simple scaling relationships to extract the effect of screening on the particle interaction energies.  By definition, the non-interacting single particle energy will not directly change with screening and thus we may write $\Delta$E$_{Opt}$ = $\Delta$E$_{BGR}$ - $\Delta$E$_{Bind}$.  Further E$_{BGR}$ = E$_{BGR}^{\epsilon_{2}=1}$/$\epsilon_{2}$ and E$_{Bind}$ = E$_{Bind}^{\epsilon_{2}=1}$/$\epsilon_{2}^{1.2}$.  In our experiment, $\epsilon_{1}$$\sim$4 for graphite \cite{Taft,Pedersen2}, initial $\epsilon_{2}$=1 (dry N$_{2}$), and final $\epsilon_{2}$=1.33$^{2}$=1.78 (in water).  Having directly measured $\Delta$E$_{Opt}$ and by using an unscreened exciton binding energy of 580 meV for nanotube (12,4) \cite{Capaz3} we are able to extract values for the screened exciton binding energy and for the screened and unscreened BGR energies.  Note that this calculated E$_{11}$ exciton binding energy is expected to be a slightly smaller\cite{Jiang} than that at E$_{22}$, so the derived values are conservative.  Specifically, this analysis yields an exciton binding energy of $\sim$290 meV after immersion in water, an unscreened BGR energy of $\sim$730 meV, and a screened BGR energy of $\sim$410 meV.  Thus large reductions in the exciton binding energy and BGR energy of $\sim$290 meV and $\sim$320 meV, respectively, lead to the small 30 meV red shift measured in the optical transition energy, depicted schematically in Figure \ref{fig3}.  Limitations of our model include: the assumption of solid wires rather than cylindrical shells in both quantum wire models; published values of $\epsilon$ are used as fit parameters when applied to heterogeneous environments; and the value of $\epsilon_{2}$ is not known precisely.

We can also compare the exciton binding and BGR energies at $\epsilon_{2}$=1 for these nanotubes.  Theory predicts that the BGR energy should be larger than the exciton binding energy which is used to explain the so-called ``ratio problem'' \cite{Kane1, Kane2, Ando}.  Indeed, we find E$_{BGR}^{\epsilon2=1}$ - E$_{Bind}^{\epsilon2=1}\sim$ +150 meV, a BGR energy larger than the exciton binding energy at $\epsilon_{2}$=1 by about 25 percent for this particular nanotube in an unscreened environment.  Qualitatively, this result is fairly insensitive to the choice of $\alpha$, going as low as +70 meV at $\alpha$=1 or as high as +220 meV at $\alpha$=1.4 at $\epsilon_{2}^{Final}$=1.78. 

Although E$_{BGR}^{\epsilon2=1}$ is greater than  E$_{Bind}^{\epsilon2=1}$, the change in E$_{BGR}$ is not necessarily larger than the change in E$_{Bind}$ with small screening by virtue of their different scaling exponents.  In fact, our model predicts that values of $\alpha$ greater than 1.4 lead to negligible or even blue shifts with small screening; the initial red shift measured in high humidity N$_{2}$ indicates a monotonic decrease in the electronic exciton energy level with increasing screening and thus supports a value of $\alpha$ less than 1.4.  Table \ref{tab2} shows the underlying variation of the particle interaction energies predicted by the model as a function of the external dielectric value, the scaling exponent, and optical transition energy shifts.  CNT 2 in Table \ref{tab1} is assigned as a (13,2) nanotube.  It belongs to the same branch (2n+m) and family as the (12,4) nanotube; thus the particle interaction energies determined by the model are very similar.

In summary, we experimentally show a monotonic decrease in the optical transition energy with increasing dielectric environment and derive a scaling relation between the exciton binding energy and external dielectric.  Our model explains the small shifts despite large changes in the underlying particle interaction energies.  Further, we demonstrate that the band gap renormalization energy is significantly larger than the exciton binding energy at $\epsilon$=1.

This work was supported by Air Force Office of Scientific Research under Grant No. MURI F-49620-03-1-0379, by NSF under Grant No. NIRT ECS-0210752, and by a Boston University SPRInG grant.

\bibliography{em}

\begin{thebibliography}{40}
\bibitem[{Sin()}]{SingleCNTs}
\bibinfo{note}{The measured nanotubes are presumed to be individual based on
  several observations: (i) the low CNT density (SEM), (ii) the measured values
  of E$_{22}$ which are very close to published values for individual CNTs, and
  (iii) no indications of resonant modes from other CNTs were found over the
  entire tunable range.}
\end{thebibliography}

\end{document}